# How Real are Liquid Groundstates? Ultra-Fast Crystal Growth and the Susceptibility of Energy Minima in Liquids

Gang Sun and Peter Harrowell[*]

*School of Chemistry, University of Sydney, Sydney New South Wales 2006 Australia*

* Corresponding author peter.harrowell@sydney.edu.au

Abstract

We calculate the degree to which the final structure of the local groundstate in a liquid is a function of the strength of a perturbing potential applied during energy minimization. This structural susceptibility is shown to correlate well with the observed tendency of liquid adjacent to a crystal interface to exhibit a crystalline groundstate, a feature that has been strongly linked to the observation of ultra-fast crystal growth in pure metals and ionic melts. It is shown that the structural susceptibility increases dramatically as the interaction potential between atoms is softened.

## 1. Introduction

In 1982 [1], Stillinger and Weber suggested that the local minima of the potential energy of a dense liquid represented a useful set of reference states with which to describe the liquid. These local minima, eventually christened 'inherent structures', have proven useful in a variety of applications in liquids and amorphous solids. They inspired the study of normal mode treatments of liquid dynamics [2], have been used to calculate configurational entropies [3], provide the basis for structural studies of amorphous materials [4] and for stochastic kinetic treatments of structural relaxation [5]. The existence and utility of inherent structures as a general concept is not in doubt. What is not clear is the 'reality' of the mapping of a liquid configuration to a *specific* inherent structure. The issue is the complexity of the energy minimization process required to realise this mapping and its possible susceptibility to differences in algorithm (e.g. steepest descent vs conjugate gradient) and small perturbations (boundary conditions, potential truncations, etc). In most applications, this issue is of little interest since, as long as the minimization process is reproducible, the exact identity of the disordered groundstate is not important. This situation changes, however, when one of the accessible groundtstates is crystalline. In this case, whether a particular minimization does or does not find an ordered inherent structure is a matter of some significance. While crystalline inherent structures have not been observed in liquids in 3D, they have been observed, as we outline below, in liquids adjacent to a crystalline solid in a crystal-liquid interface [6].

This story starts with the observation that a number of pure metals can achieve extraordinary crystal growth rates from the supercooled melt. In nickel, for example, this rate can reach a value of 70 m/s [7]. This corresponds to the formation of a new crystal layer every 5 picoseconds, a time interval that corresponds to only 5 periods of oscillation at the Debye frequency. Furthermore, this rapid growth rate shows no sign of activated control [8], a kinetic feature typically associated with cooperative ordering. This extraordinary kinetic behavior appears to be restricted to the face centred crystal (for pure metals) and the NaCl structure (for ionic melts) [9]. In 2018 [6] we identified the origin of these extreme growth rates by showing that the liquid side of the crystal-liquid interface of these ultra-fast crystallizers had local potential energy minima that were crystalline. This observation is in sharp contrast with the inherent structures in the bulk liquid, the so-called inherent structures [10], which are always disordered in 3D. With an ordered groundstate, the crystal growth rate is no longer governed by the rate of cooperative reorganization of the liquid configuration.

Instead, the growth rate is limited only by the rate at which the vibrational energy can be removed, a process without any activation barrier.

The demonstration of the existence of the crystalline inherent structures in the interface solved the problem of the fast barrierless crystal growth in metals and salts but it did so by shifting the puzzle to the origin of these order inherent structures. After all. if the liquid side of the crystal-melt interface possess a 'hidden' order in the form of a crystalline groundstate that extends out beyond the crystal interface, wouldn't the formation of this groundstate, itself, require some sort of cooperative rearrangement? And if so, wouldn't this hidden propagation involve a barrier of some sort which would then be the rate determining step in growth? The fact that the steady state growth rate of a metal like nickel shows no sign of such activation control leaves us with two options: either the propagation of the crystalline inherent structure is achieved by some still unknown barrierless process, or there is actually no hidden ordering process at all. In this paper we shall present evidence in support of this second option and explore how the details of the interparticle potential influences the appearance crystalline order in the interfacial inherent structures.

How could crystalline order appear in the interfacial inherent structure without any underlying ordering process occurring? The key to the answer is to recognise that an inherent structure is defined by an energy minimization procedure, a nontrivial descent through a complex energy landscape. It is plausible that a perturbation during this minimization, such as that provided by the adjacent crystal interface, could divert the minimization, guiding it to a nearby (in configuration space) crystalline minimum. Under this perturbation, an initial configuration that is no different from that of a bulk liquid can end up at a crystalline local energy minimum. For this proposition to be validated it is necessary to demonstrate that the energy minimization of an equilibrium configuration of a bulk liquid can result in a crystalline groundstate when subjected to a weak perturbing field.

The logic of this paper is as follows. First, we test the proposition that the minimization process in a liquid is structurally susceptible by applying a template field to determine the critical strength λ* of the applied field is required to generate order during minimization, i.e. to locate a crystalline inherent structure. Then, we identify this critical field strength $\lambda^*$ for a variety of liquids and compare the trend in $\lambda^*$ obtained from the *bulk* liquid with the observation of crystalline inherent structures in *interfacial* liquid. Here we make the tacit assumption that the response of the liquid to the inhomogeneous ordering field generated by the interface is similar to the response to a homogeneous field. Finally, we confirm that the presence of crystalline interfacial inherent structures correlates strongly with the low activation barriers for crystal growth, essentially redoing the analysis of ref. [6] for the liquid models studied here.

## 2. Model and Algorithm

The model liquid studied in this paper is a pure atomic liquid interacting via a Morse potential,

$$\phi(r) = \varepsilon \left\{ \exp\left[-2\gamma\left(\frac{r}{\sigma}-1\right)\right] - 2\exp\left[-\gamma\left(\frac{r}{\sigma}-1\right)\right] \right\} \quad (1)$$

with ε = 0.3429 eV, γ = 3.894 and σ = 2.866 Å. These parameters have been chosen to reproduce the equation of state and elastic constants in copper [11]. All liquids are simulated at a fixed pressure of 20k bar and N = 32000, unless otherwise indicated, using the LAMMPS molecular dynamics algorithm [12]. Potential energy minimization is carried out using a conjugate gradient algorithm at fixed volume. The simulation cell for the energy minimization calculations was 77×77×77 (Å$^3$) in a shape of cube with periodic boundary conditions applied.

We shall carry out energy minimizations in the presence of an applied potential field U({r}). This potential, which is applied throughout the simulation volume, is 'turned on' only at the start of the minimization. Once the minimum with the applied field U has been reached, we then turn off the field and continue the minimization so that the final minima are true inherent structures of the unperturbed liquid potential energy. Note this potential is applied *only* during the minimization so the initial liquid configuration is a representative configuration of the unperturbed equilibrium liquid state. In this study we have chosen U to be

$$U(\{r_i\}) = \lambda \sum_{i}^{N} \sum_{m}^{N} \phi(|\vec{r}_i - \vec{R}_m|) \qquad (2)$$

where $\phi(r)$ is the same as the pairwise potentials (see Eq.1) and $\vec{R}_m$ is the $m$th lattice site of an FCC crystal, corresponding to that of a perfect crystal equilibrated at the same pressure as the liquid and at a temperature equal to the melting point and extended throughout the simulation cell. The melting temperatures for the Cu model (i.e. $\gamma$ = 3.894) is 2250K. The analogous melting points when $\gamma$ is set to 8.598 and 17.196 are, respectively, 2644K and 2500K. The parameter $\lambda$ is varied to control the magnitude of the templating potential and the resulting degree of crystalline order in the final configuration measured using average value of $\bar{Q}_6$ which is the average form of the local bond order parameter $Q_{6,i}$ over all its neighbors and itself [13]. (The quantity $Q_{6,i}$ is the $l$ = 6 spherical harmonic of the nearest neighbour density of the $i$th particle.)

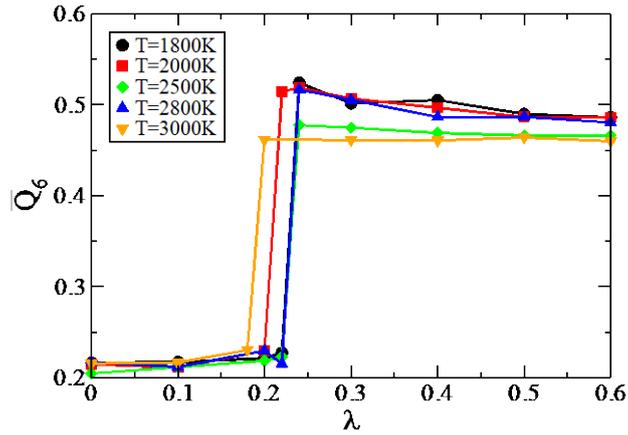

**Figure 1.** The crystalline order parameter $\overline{Q}_6$ of the potential energy minimum plotted as a function of λ, the amplitude of the templating potential U (see Eq. 2).

## 3. Measuring the Susceptibility of the Minimization Process

In Fig. 1 we plot the crystalline order parameter $\overline{Q}_6$ of the potential energy minimum plotted as a function of λ, the amplitude of the templating potential. We see an abrupt increase in crystalline order in the solid at a field with an amplitude λ ~ 0.2, i.e. roughly 20% of the amplitude of the individual particle interactions. Details of the spatial character of the ordering during templated minimization are provided in the Appendix. This threshold value of λ shows little dependence on the temperature of the initial configuration. As the λ = 0 result shows, the inherent structures of the equilibrium liquid exhibit little in the way of crystal-like order. It appears that minimization is stable with respect to the perturbing field up to the point that the field sufficiently distorts the energy surface so as to open a barrierless path to crystalline order. Note that this ordering is achieved through the perturbed minimization alone. The results of Fig. 1 establish a major result of this paper, namely that the local groundstate (or inherent structure) of a liquid configuration is not an immutable property of that configuration but is susceptible, in this case dramatically so, to a perturbation of the minimization process by which it is generated. As the energy minimization and thermal cooling are analogous processes, the nonlinear susceptibility of the inherent structure is of direct significance in the kinetics of liquid ordering.

## 4. Field-Guided Inherent Structures and Ultra-Fast Crystal Growth

Having established the susceptibility of the inherent structure to a field applied during minimization, our next task is to understand how this susceptibility depends on the interaction potential between atoms. To determine how the susceptibility of the inherent structure varies across a series of liquids, we shall repeat the analysis of Fig. 1 for liquids with different values of the softness parameter $\gamma$ of the Morse potential. As shown in Fig. 2a, the effect of increasing $\gamma$ is to stiffen the short range repulsion and narrow the width of the attractive well. The liquid pair correlation function, g(r), exhibits an increasingly sharply defined first peak as $\gamma$ increases, as shown in Fig. 2b. We have held the temperature (T = 3000K) and the pressure (p = 20k bars) fixed for all values of $\gamma$.

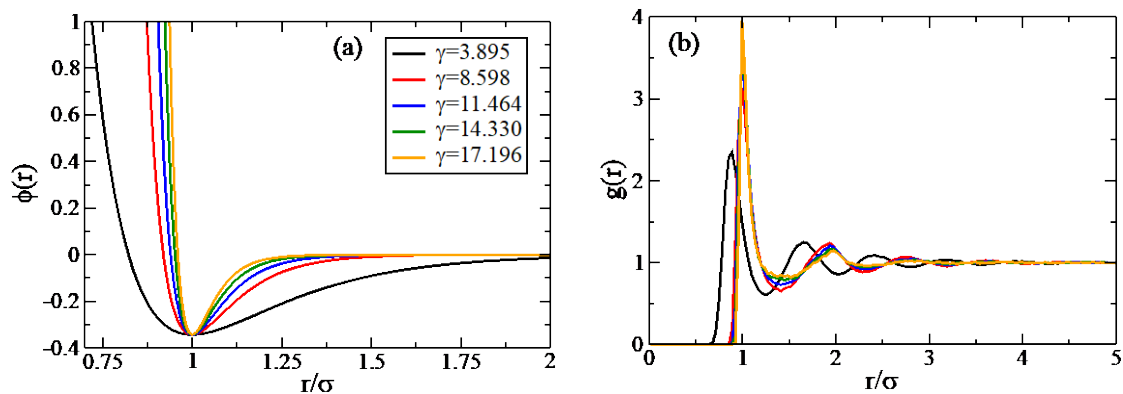

**Figure 2**. a) Plots of the interaction potentials $\phi(r)$ vs r/$\sigma$ for different values of $\gamma$ as indicated. b) The pair distribution function g(r) for the different Morse liquids, calculated at T = 3000K and a pressure of 20k bars.

We have repeated the perturbed minimization calculations as described previously. The value of the degree of crystallinity in the inherent structure as a function of $\gamma$ is plotted in Fig. 3.

We find a marked drop in the susceptibility of the liquid to the influence of the perturbing field during minimization as γ increases from 9 to 11. By the time we get to γ = 17.196 we find little evidence of susceptibility remains. These results demonstrate a clear qualitative difference between 'soft' liquids (i.e. γ < 9) and 'stiff' liquids with larger γ. For context, most metals and ions are 'soft' as can be seen from the values of γ used in empirical Morse potentials, i.e. 3.184 (Na), 3.788 (Al), 3.170 (K), 3.975 (Fe), 3.947 (Ni), 4.264 (Ag) and 4.280 (W), as reported in ref. [11].

Note that the change in the value of γ changes both the interaction potential (Eq.1) and the applied field U (Eq. 2). This means that as we increase the stiffness of the liquid interactions (i.e. increase γ) we also increase the stiffness of the perturbing field. So, should the increase in the critical field strength $\lambda^*$ due to the increase in γ observed in Fig. 3 be associated with the change in the atomic interactions or the perturbing field? We maintain that the reduced susceptibility observed on increasing γ is a property of the liquid, not the perturbing potential. The argument is simply that increasing γ of the perturbing field will always increase the degree of ordering induced in the liquid. Since we observe a *decrease* in ordering with increasing γ it follows that this must be due to the changing stability associated with the pairwise interactions.

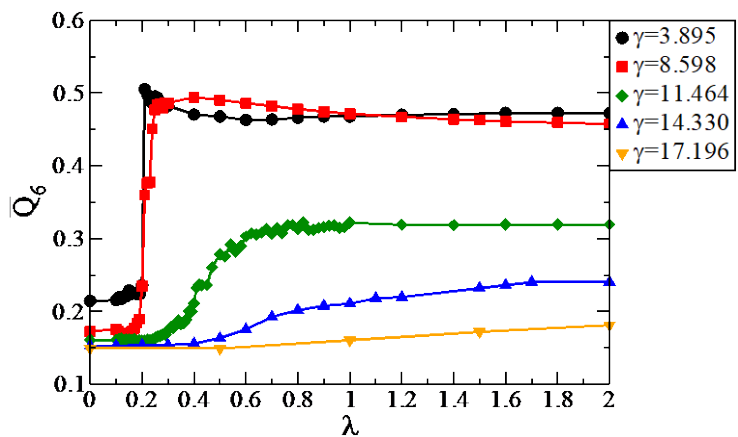

**Figure 3.** The crystalline order parameter $\bar{Q}_6$ of the potential energy minimum for liquids with a range of values of γ plotted as a function of λ, the amplitude of the templating potential U. The initial liquid configurations were equilibrated at T=3000K in all cases.

Our contention is that the occurrence of crystalline inherent structure in the liquid adjacent to the crystal-liquid interface is a consequence of the susceptibility of these inherent structures to the effective field exerted by the adjacent crystal. Having established that the susceptibility increases strongly as γ decreases below a value of ~ 9, we predict that we should find an increasing degree of crystalline order crystalline inherent structures for 'soft' liquids and only disordered interfacial inherent structures for liquids with γ > 9. We verify this prediction in Fig. 4, where we plot the crystal-liquid interface before and after minimization. In the case of γ = 8.598, the interface advances when minimized, explicit evidence of crystalline inherent structure. In the case of γ = 17.196, there is no sign of an ordered interfacial inherent structure. These results are consistent with the variation in structural susceptibility presented in Fig. 3 and so support the proposal that influence of the interface acts similarly to the applied field U.

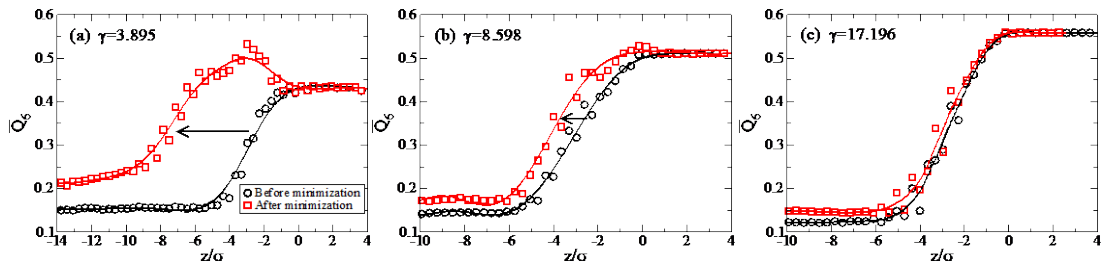

**Figure 4.** The (111) crystal liquid interface measured as $\bar{Q}_6$ vs z/σ, the normal distance, before (black) and after (red) energy minimization for a) γ = 3.895, b) γ = 8.598 and c) γ = 17.196, as indicated. The advance of the interface during minimization indicates the crystalline order in the interfacial inherent structure.

The final link in our argument is to establish the correlation between the crystalline interfacial inherent structures and ultra-fast crystal growth. In ref. 4, we did this for liquids interacting via EAM potentials and here we shall do the same for the Morse liquids. In Fig. 5a we plot the temperature dependence of the crystal growth rate V as a function of temperature. Each plot is truncated at a low T due to the kinetic instability of the supercooled liquid with respect to crystallization. As shown previously [6], the spontaneous formation of randomly oriented crystallites impedes the propagation of the crystal front, causing the observed growth rate to turnover and decrease with further cooling and, in so doing, rendering the growth rate ill-defined.

Over the accessible temperature range, the ordering kinetics at the interface is conveniently quantified in terms of the rate coefficient for crystal addition, $k$, which can be obtained from the steady state growth rate V(T) via the relation,

$$V \approx k \frac{\Delta h}{k_B T} \frac{(T_m - T)}{T_m} \quad , \tag{3}$$

the expansion of the Wilson-Frenkel expression for V(T) [15] to first order in $\frac{\Delta h}{k_B T} \frac{(T_m - T)}{T_m}$.

Here $T_m$ is the melting point and $\Delta h_m$ is the enthalpy of fusion per particle at $T_m$. The fit of Eq.3 to the simulated growth rates in Fig. 5a were achieved by treating the coefficient k as a constant, independent of T. In Fig. 5b, we plot these fitted values of k against the interaction stiffness γ and note that the sharp increase in structural susceptibility is matched by an analogously steep increase in the attachment rate. In ref. [6] we considered how the presence of a crystalline inherent structure at the interface would influence to nature of the coefficient k and its temperature dependence. In ref. [9], we reported examples of crystal growth in which the absence of a crystalline interfacial inherent structure as associated with activated growth kinetics, i.e. an attachment coefficient $k(T) \sim \exp[-\beta E_a]$. We have not found that to be the case in data for the Morse liquids. The attachment coefficient k for the $\gamma = 17.196$ liquid shows no significant temperature dependence, over the accessible range of supercoolings, despite the absence of any order in the interfacial groundstate.

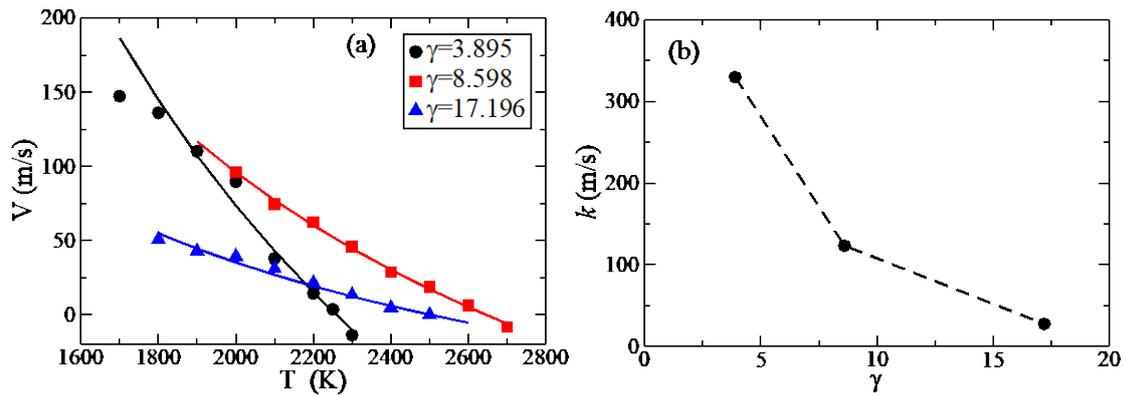

**Figure. 5** a) Crystal growth rates V vs temperature for three values of $\gamma$, as indicated. The data is truncated at the lowest temperature before the onset of the kinetic spinodal instability of the liquid (see text). The solid curves are fits to Eq. 3.  b) The attachment coefficient k obtained from the fits in Fig. 5a plotted against $\gamma$. Note the rapid increase as $\gamma$ decreases below 9.

## 5. Conclusions

In this paper we have demonstrated that the ordered inherent structures of the liquid adjacent to the crystal interface, the feature responsible for the extremely high rate of particle attachment observed during crystal growth in metals and salts, do not necessarily arise from any sort of ordering, hidden or overt, of the equilibrium structure in the interfacial liquid. Instead, we have shown that, for particles with sufficiently soft interactions, the liquid can be induced to order during minimization due to the influence of a weak perturbing field of the adjacent crystal. This result allows us to complete the explanation of fast crystal growth initiated in ref. [6] by demonstrating that the melts of metals and atomic ions are generally so structurally susceptible that the presence of a crystal interface could be sufficient to guide them to the crystal as energy is rapidly removed without the need for the stochastic cooperative search of configuration space traditionally assumed to be necessary. These results do not exclude the possibility that the crystal perturbs the structure of the interfacial liquid at equilibrium. What we have demonstrated is that for a class of liquids, here identified as having soft interactions as measured by $\gamma < 9$, such equilibrium structural perturbations are not necessary for the existence of a crystalline interfacial inherent structures and, hence, small activation barriers to growth.

The demonstration of inherent structural susceptibility has potential applications beyond the explanation of rapid crystal growth. We can ask, for example, about what other structures populate the space of a liquid's inherent structures and so can be similarly accessed by a perturbing potential. These other structures might include metastable polymorphs and specific disordered glasses. The results in the paper also raise the possibility that a metal or molten salt might be used as rapid memory storage device, where the information to be stored is in the form of a specific template structure.

**Acknowledgements**

The authors acknowledge the support of the Australian Research Council through the Discovery Grant program.

**Data Availability**

The data that support the findings of this study are available from the corresponding author upon reasonable request.

**Appendix. The Nature of Crystal Ordering during Templated Minimization.**

In Fig. 1 we report the volume average value $\bar{Q}_6$ of the local crystalline order parameter $Q_{6,i}$ at the end of the energy minimization during which the crystal template potential is applied. To provide a more detailed picture of just how this ordering takes place we can examine the evolution of the distribution of the local values $Q_{6,i}$ for progressive steps through the minimization. In Fig. 6 below we plot the distribution of values of $Q_{6,i}$ after n steps of the minimization procedure. The minimization in question is the mode Cu liquid with a template field strength of $\lambda = 0.6$. We see that the order evolves about a roughly bimodal distribution with peak, in the initial liquid, of $\bar{Q}_6 = 0.12$, and a peak corresponding to a crystal environment peaked at $\bar{Q}_6 = 0.55$. Direct imaging of the spatial distribution of crystal order during the minimization is provided in Fig. 7.

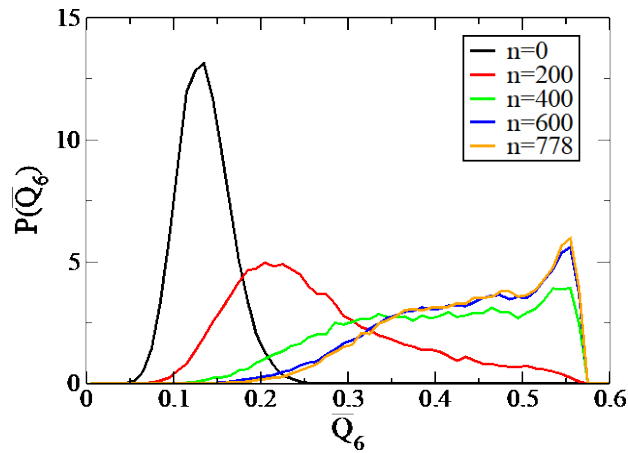

**Figure 6.** The distribution of the local order parameter $\bar{Q}_6$ during the templated minimization of Cu with $\lambda = 0.6$. The curves correspond to the structure distributions after n steps of the minimization, as indicated.

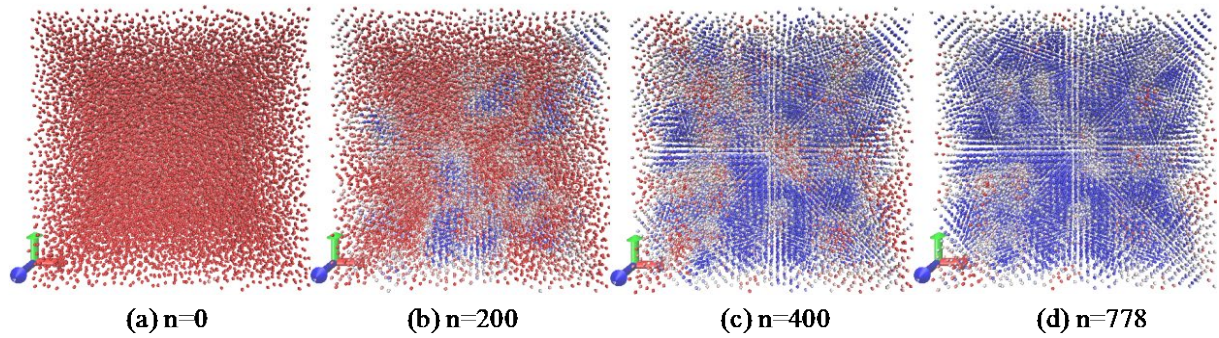

(a) n=0  (b) n=200  (c) n=400  (d) n=778

**Figure 7**. Images of the spatial distribution of crystalline order $\bar{Q}_6$ during the templated minimization as indicated by the number of minimization steps referred to in Fig. 6. The color scheme is graded between red (low $\bar{Q}_6$) to blue (high $\bar{Q}_6$). Note the localized patches marking the initial appearance of crystal structure (n = 200) and their subsequent growth.

We find that ordering during minimization occurs heterogeneously, starting in local independent patches (see n = 200 in Fig. 7) which then grow out as the minimization algorithm precedes. This result demonstrates that under the influence of uniform crystal field, it is local patches of the liquid that are initially 'steered' towards crystallinity. While this process is reminiscent of crystal nucleation, it is distinct in that the system here evolves via irreversible minimization and so there is no question of crystal clusters having to exceed some critical size as is the case in nucleation at metastable equilibrium.

**References**

1. F. H. Stillinger and T. A Weber, Hidden structure in liquids. *Phys. Rev. A* **25**, 978-989 (1982).

2. R. Zwanzig, On the relation between self-diffusion and viscosity in liquids. *J. Chem. Phys.* **79**, 4507-4508 ((1983).

3. F. Sciortino, W. Kob and P. Tartaglia, Inherent structure entropy of supercooled liquids. *Phys. Rev. Lett.* **83**, 3214-3217 (1999).

4. C. P. Royall and S. R Williams, The role of structure in dynamical arrest. *Phys. Rep.* **560**, 1-75 (2015).

5. A. Heuer, Exploring the potential energy landscape of glass-forming system: from inherent structures via metabasins to macroscopic transport. *J. Phys.: Cond. Matt.* **20**, 373101 (2008).

6. G. Sun, J. Xu and P. Harrowell, The mechanism of the ultrafast crystal growth of pure metals from their melts, *Nature Mater.* **17**, 881 (2018).

7. O. Funke, G. Phanikumar, P. K. Galenko, L. Chernova, S. Reutzel, M. Kolbe and D. M. Herlach, Dendrite growth velocity in levitated undercooled nickel melts, *J. Cryst. Growth* **297**, 211-222 (2006).

8. Coriell, S. R. and Turnbull, D. Relative roles of heat transport and interface rearrangement rates in the rapid growth of crystals in undercooled melts. *Acta Metall.* **30**, 2135–2139 (1982).

9. A. Hawken, G. Sun and P. Harrowell, Role of interfacial inherent structures in the fast crystal growth from molten salts and metals, *Phys. Rev. Mater.* **3**, 043401 (2019).

10. C. P. Royall and S. R. Williams, The role of local structure in dynamical arrest, *Phys. Rep.* **560**, 1-75 (2015).

11. L. A. Girifalco and V. G. Weizer, Application of the Morse Potential Function to Cubic Metals, *Phys. Rev.* **114**, 687 (1959).

12. S. J. Plimpton, Fast parallel algorithms for short-range molecular dynamics. *J. Comp. Phys.* **117**, 1–19 (1995).

13. W. Mickel, S. C. Kapfer, G. E. Schröder-Turk and K. Mecke, Shortcomings of the bond orientational order parameters for the analysis of disordered particulate matter. *J. Chem. Phys.* **138**, 044501 (2013).

14. E. Burke, J. Q. Broughton and G. H. Gilmer, Crystallization of fcc (111) and (100) crystal-melt interfaces: A comparison by molecular dynamics for the Lennard-Jones system. *J. Chem. Phys.* **89**, 1030-1041 (1988).

15. K. A. Jackson, K. A. The Interface Kinetics of Crystal Growth Processes. *Interface Sci.* **10**, 159–169 (2002).